\documentclass[useamsfonts]{pasj00}

\SetVolumeData{2006}{58}{1}
\renewcommand{\vec}[1]{\mbox{\protect\boldmath$#1$}}
\newcommand{\rs}{R_{\mbox{\tiny S}}}
\newcommand{\rin}{R_{\rm in}}
\newcommand{\rout}{R_{\rm out}}
\newcommand{\rph}{R_{\rm ph}}
\newcommand{\risco}{R_{\mbox{\tiny ISCO}}}

\newcommand{\el}{\mathrm{e}}
\newcommand{\st}{\sigma_{\mbox{\tiny T}}}

\newcommand{\bge}{\langle\gamma_\mathrm{e}\rangle}
\newcommand{\dd}{\mathrm{d}}
\newcommand{\der}[2]{\frac{\,\dd #1}{\,\dd #2}}

\usepackage{graphicx}
\begin{document}
\hyphenation{Schwarz-schild}

\SetRunningHead{J.~Hor\'ak \& V.~Karas}{Polarization of light from warm clouds above an accretion disk}
\Received{2005/11/17}
\Accepted{2005/12/05}

\title{Polarization of light from warm clouds above an accretion disk:\\
       effects of strong-gravity near a black hole}

\author{Ji\v{r}\'{\i}~\textsc{Hor\'ak} and Vladim\'{\i}r~\textsc{Karas}}
\affil{Astronomical Institute, Academy of Sciences, Bo\v{c}n\'{\i}~II, CZ-141\,31~Prague, Czech~Republic}
\maketitle
\begin{abstract}
We study polarization from scattering of light on a cloud in radial
motion along the symmetry axis of an accretion disk. Radiation drag from
the disk and gravitational attraction of the central black hole are
taken into account, as well as the effect of the cloud cooling in the
radiation field. This provides us with a self-consistent toy-model for
predicted lightcurves, including the linear polarization that arises
from the scattering. Strong gravitational lensing creates indirect
images; these are formed by photons that originate from the disk, get
backscattered onto the photon circular orbit and eventually redirected
towards an observer. Under suitable geometrical conditions the indirect
photons may visibly influence the resulting magnitude of polarization
and light-curve profiles. Relevant targets are black holes in active
galactic nuclei and stellar-mass Galactic black-holes exhibiting
episodic accretion/ejection events.
\end{abstract}

\section{Introduction}
\label{intro}
The present-day evidence for black holes relies almost entirely on
information carried by electromagnetic waves. X-rays play a particular
role (\cite{sew95}): they are supposed to emerge from gas near a black
hole horizon and bring us imprints of physical processes and conditions
in the place of their origin and along the ray path. Various spectral
and lightcurve patterns have been identified as likely signatures of
supermassive black holes in galactic nuclei and stellar-mass black
holes. These features presumably arise when matter is accreted from an
immediate vicinity of the black hole. The ultimate goal of this 
effort is to prove the existence of event horizons in `real' nature and
this way to discover black holes. Clearly, an affirmative proof is a
great challenge that may still be far ahead of us. Here we study a
related task, which appears to be somewhat easier on the technical
side, though it also represents an unresolved issue as yet: searching for
rays of photons that encircle an ultra-compact star or a black hole.

Accretion disks represent a common way of feeding black holes and
generating photons, which then experience strong-gravity while
travelling to a distant observer (\cite{kat98};
\cite{kro99}). Outside the disk plane density is much less, however,
even regions near axis are not empty: winds and jets emerge, roughly
along the symmetry axis. We address a question whether future
observations of polarized time-dependent signal can help recognizing the
effects of strong-gravity light bending, and what features are expected
in light curves. To this aim we examine a model of warm clouds moving
radially along the symmetry axis. Primary photons from the disk are
Thomson-scattered and polarized. We assume that the process takes place
near a Schwarzschild black hole, where the higher-order (indirect) light
rays contribute to the observed  radiation flux. The resulting
modulation of intensity and polarization magnitude can reach a
non-negligible level under suitable geometrical alignment of the black
hole, the cloud and the observer. We treat the interaction of the cloud 
with the radiation and gravitational fields in the relativistic
framework (\cite{abr90}; \cite{vok91};
\cite{kea01}). This approach provides us with a self-consistent
description of the cloud motion and the resulting observed signal. Both
ingredients are conveniently expressed in terms of the radiation
stress tensor.

There is the evidence for jets being formed only a few tens
gravitational radii from a supermassive central black hole
(\cite{jun99}). The emission mechanisms producing the observed
high-energy photons (X- and $\gamma$-rays) are likely non-thermal, but
it is not clear whether the synchrotron emission or the inverse Compton
emission dominates in each particular source (\cite{har02}). Here we
concentrate on the latter case (within the Thomson approximation), which
seems to be relevant for radio-quiet sources with the ambient radiation
acting on particles and fluids and determining their terminal
(equilibrium) speed; see \cite{noe74}; \cite{ode81}; \cite{sik81}; and
\cite{phi82} for original papers. This dynamical influence of the
radiation field has been studied further by Sikora et al.\
(\yearcite{sik96}), Renaud \& Henri (\yearcite{ren98}), Tajima \& Fukue
(\yearcite{taj98}), Fukue (\yearcite{fuk99}) and Watarai \& Fukue
(\yearcite{wat99}). Ghisellini et al.\ (\yearcite{ghi04}) proposed a
model of aborted (failed) jets in which colliding clouds and shells
occur very near a black hole and are embedded in strong radiation.
According to their scheme, most of energy dissipation should take place
on the symmetry axis of an accretion disc where the individual clouds
either move away from the black hole, or they fall back. The
process of gravitational and radiative acceleration of plasma was
studied also by Fukue et al.\ (\yearcite{fuk01}) and Fukue
(\yearcite{fuk05}), who examined the efficiency of collimation towards
the disk axis and applied this model to the case of microquasars.

The effects of strong gravity are significant in this region and the
challenge for future techniques is to identify subtle, yet specific
patterns in X-ray lightcurves. Polarization studies could help to
achieve this goal. Distinct features should arise from multiple
trajectories of light rays connecting the source with the observer along
several different paths. The rays winding up around the photon circular
orbit should experience a characteristic mutual time delay. We discuss
the expected features in this paper.

Polarization of light from scattering in winds and jets was examined by
various authors. Following the early papers (e.g.\ \cite{dol67};
\cite{ang69}; \cite{bon70}), Begelman \& Sikora (\yearcite{beg87})
studied the linear polarization of initially unpolarized soft radiation 
up-scattered by cold electrons in a jet. Beloborodov (\yearcite{bel98})
examined the case of fast winds outflowing from an accretion disk slab.
He found that polarization direction depends on the wind velocity; the
terminal speed of the outflow plays a critical role. Poutanen
(\yearcite{pou94}) and Celotti \& Matt (\yearcite{cel94}) considered the
synchrotron self-Compton mechanism, a likely process producing
polarization wherever magnetic fields interact with relativistic
particles. The effect of electron temperature was also discussed: it
reduces the final magnitude of polarization. Recently, Lazzatti et al.\
(\yearcite{laz04}) further discussed the Compton drag as a conceivable
mechanism for polarization in gamma-ray bursts. Hor\'ak \& Karas
(\yearcite{hor05}) studied the polarization of scattered light from a
compact star, taking into account the light-bending effect. In fact, it
was demonstrated that retro-lensing images, which clearly require strong
gravity, can give rise to specific polarimetric signatures in predicted
lightcurves. Here we develop this model further by considering
non-negligible temperature of the scattering medium and by changing the
geometry of the primary source. We assume a Keplerian disk as the
source, so we are able to examine situations that are relevant for
accreting black holes.

\begin{figure}[tb!]
\begin{center}
\FigureFile(0.48\textwidth,0.48\textwidth){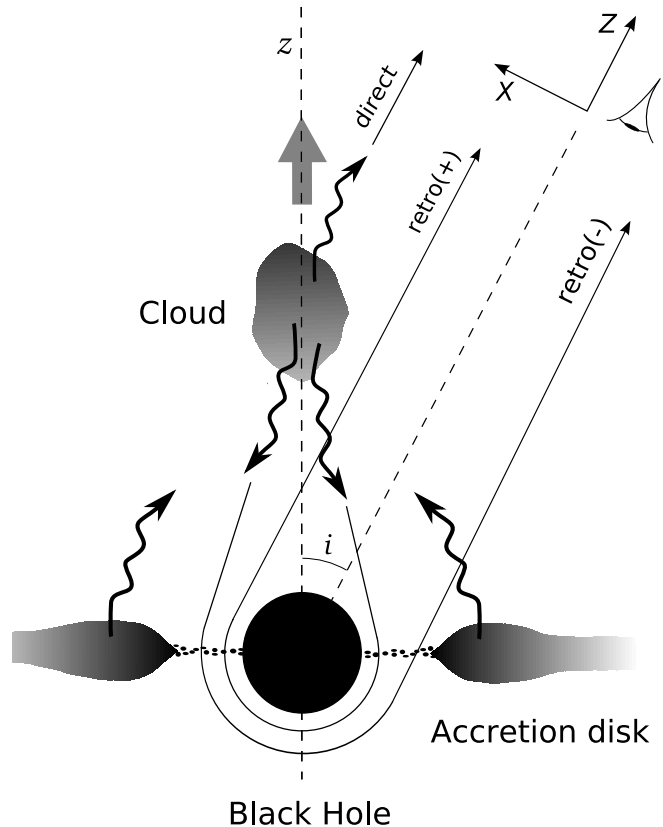}
\end{center}
\caption{Geometry of the model. An accretion disk is the source of primary
unpolarized light, which is then Thomson scattered on a cloud. The cloud 
moves radially along the axis, $z\,\equiv\,z(t)$, in the radiation field
of the disk. Gravity of the central black hole influences the motion of the
cloud. The light rays of primary and scattered photons are also affected. 
Direct and indirect (retro-lensed)
light rays have different degree of linear polarization and they
experience different amplification and the Doppler boosting. The observed
lightcurve is produced by all the rays reaching an observer at view angle
$i$ far from the centre (along
$Z$-direction). The indirect photons contribute  most significantly to the
the total signal if the cloud moves toward the black hole and the
observer inclination is small.}
\label{fig1}
\end{figure}

\section{The model setup}
As a toy-model for non-uniform outflows and aborted jets we consider a
cloud of particles moves through the radiation field of a standard thin
accretion disc (\cite{sha73}). The disk defines the
equatorial plane $\theta=\pi/2$ of the system. Primary photons from the
disc are scattered by electrons in the cloud, they are beamed
(preferentially in the direction of the scatterer motion), and polarized
by Thomson mechanism. We adopt the first scattering approximation (small
optical depth $\tau\ll1$ of the cloud is assumed) and we restrict the
cloud motion to the axis of symmetry, $z\,\equiv\,z(t)$. The basic setup
of the model is captured in figure~\ref{fig1}. 

The spacetime metric is (\cite{mis73})
\begin{equation}
 \dd s^2 = -\xi \dd t^2 + \xi^{-1}\dd r^2 + r^2\left(\dd\theta^2 +
 \sin^2\theta\dd\phi^2\right),
 \label{eq:acc-ds}
\end{equation}
where $\{t,r,\theta,\phi\}$ are Schwarzschild coordinates, 
$\xi(r)\,\equiv\,1-\rs/r$ is the redshift function,
$\rs\,\equiv\,2GM/c^2$. Geometrized units $c=G=1$ will be used.

\begin{figure}[tb!]
\begin{center}
\FigureFile(0.48\textwidth,0.48\textwidth){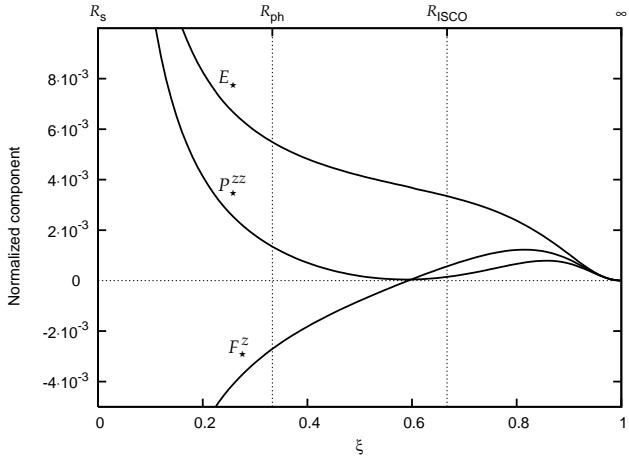}
\end{center}
\caption{Normalized components of the radiation stress tensor on the
symmetry axis for radiation originating from a standard disk. The inner
edge of the disk corresponds to the innermost stable circular orbit,
$\rin=\risco=3\rs$; no intervening matter is considered below that
radius. The outer edge is at $\rout=10^3\rs$. Location of the photon
circular orbit $\rph=1.5\rs$ is also indicated.}
\label{fig:acc-tensor}
\end{figure}

The total bolometric luminosity of a standard disk, $L=GM\dot{M}/2\rin$, 
can be scaled by the Eddington luminosity
$L_\mathrm{E}=4{\pi}cGMm_\mathrm{p}/\st$. We thus define 
$\Lambda\,\equiv\,L/L_\mathrm{E}$. The corresponding intensity emitted
from the disk is
\begin{equation}
 I_\mathrm{d}(r)=\frac{m_\mathrm{p}c^3}{\st\rs}\,\Lambda\,I_{\mathrm{d}\star},\quad
 I_{\mathrm{d}\star}\,\equiv\,\frac{3\rs^2\rin}
 {2\pi r^3}\left(1 - \sqrt{\frac{\rin}{r}}\,\right), 
 \label{ADSurfaceIntensity}
\end{equation}
where we denoted, by an asterisk, the intensity rescaled to the 
dimension-less form. Hereafter the radiation terms will be made dimension-less 
by scaling them with $(m_\mathrm{p}c^3/\st\rs)\,\Lambda$.

Photons travel along null geodesics and give rise to the radiation
stress tensor at every point outside the disk plane and above the black
hole horizon. Figure~\ref{fig:acc-tensor} shows the dependence of the
components relevant for our calculation on symmetry axis of the
black hole -- accretion disk system (we employ similar notation for 
the radiation stress tensor components as e.g.\
\cite{fuk99}; \cite{wat99}). 

In spite of the fact that we employ a particular model of a thin 
accretion disk in this paper, several rather general
features are captured in Fig.~\ref{fig:acc-tensor} that we expect 
to shape the radiation field of astrophysically realistic models too.
These can be summarized as follows.
\begin{description}
\item[--]~The radiation stress tensor vanishes for
$z\rightarrow\infty$ ($\xi\rightarrow1$) because the angular size 
of the disk, as observed by distant observers, decreases to zero.
\item[--]~Far from the centre, $tz$-component of the stress
tensor (measured by a static observer) is positive, however,
this term changes its sign at certain value of height $z$ near 
above the disk. This effect operates for a static observer inside photon 
circular orbit, who intercepts photons falling into the black hole.
\item[--]~All three components diverge 
as $z\rightarrow\rs$ ($\xi\rightarrow0$). Close to the
horizon the accretion disk appears again as a point-like source 
producing the energy density $\propto\xi^{-1}(z)$.
Here, two effects act against each
other: the gravitational bending of light-rays reduces the solid angle
that the disk occupies on the observer local sky by factor
$\propto\xi$, whereas the gravitational redshift 
increases the radiation intensity by factor $\propto\xi^{-2}$
(blueshift occurs, in fact). 
\end{description}

Intensity of scattered light and its polarization are characterized by
Stokes parameters (\cite{cha60}; \cite{ryb79}). These can be obtained by
integrating over all directions $\vec{n}_\mathrm{i}$ of photons incident
on scattering particles of the cloud from different points in the disk
(\cite{hor05}):
\begin{eqnarray}
 \bar{I} &=& \bar{E} + \bar{P}^{ZZ},
\label{eq:scattI} \\
 \bar{Q} &=& \bar{P}^{YY} - \bar{P}^{XX},
\label{eq:scattQ} \\
 \bar{U} &=& -2\bar{P}^{XY},
\label{eq:scattU}
\end{eqnarray}
where bars denote quantities in the comoving orthonormal polarization
frame, $\bar{X}$-axis of which is given by projecting the symmetry axis
onto the observing plane and $\bar{Z}$-axis coinciding with the
direction of scattered photons. $\bar{P}^{AB}$ are local components of
the radiation stress tensor (the spatial part of the energy-momentum
tensor $\bar{T}^{(\alpha)(\beta)}$), $\bar{E}\,\equiv\,\bar{T}^{(t)(t)}$
is local energy-density. The values of Stokes parameters
(\ref{eq:scattI})--(\ref{eq:scattQ}) are normalized by
$A\,\equiv\,3\tau/16\pi$, $\tau\,\equiv\,n_\el{\st}R$ is the Thomson
optical depth of the cloud in terms of electron density $n_\el$,
$R\ll\rs$ is radius of the cloud. Polarization is purely linear ($V=0$)
and its the magnitude is equal to
$\Pi=(\bar{Q}^2+\bar{U}^2)^{1/2}/\bar{I}$.

\begin{figure*}[tb!]
\begin{center}
 \FigureFile(0.48\textwidth,0.48\textwidth){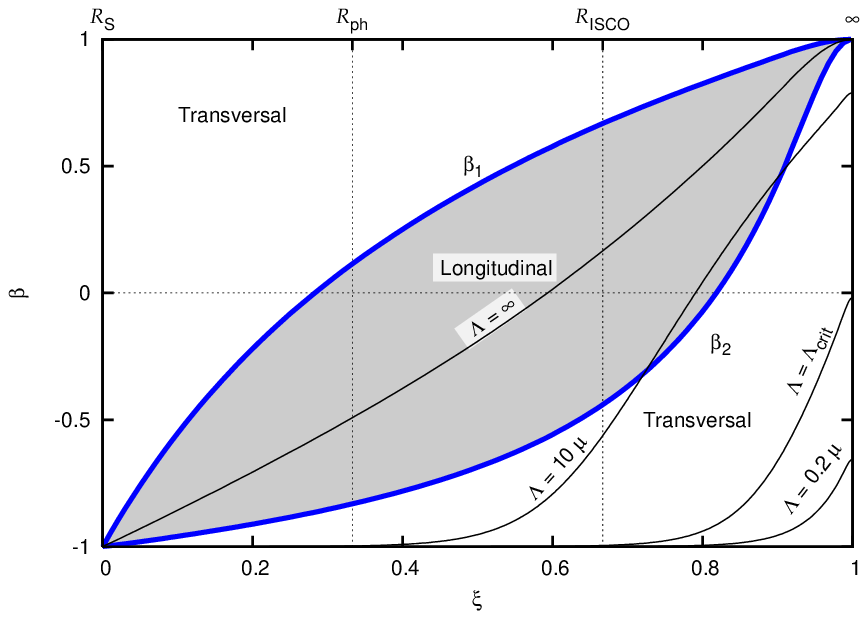}
 \hfill
 \FigureFile(0.48\textwidth,0.48\textwidth){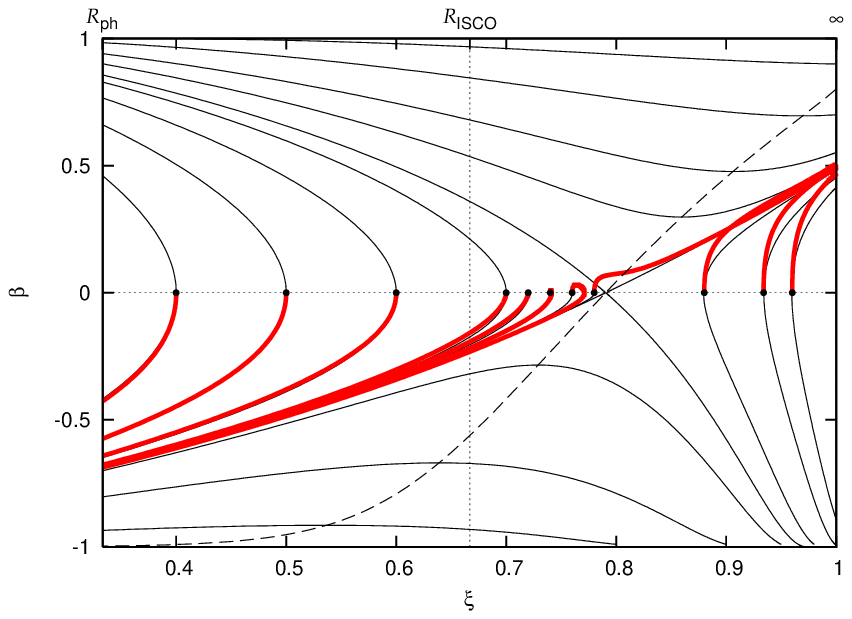}
\end{center}
\caption{Left: critical velocities of the cloud motion, $\beta_1(\xi)$
and $\beta_2(\xi)$, at which the observed polarization vector changes
its orientation between the longitudinal (inside the shaded region) and 
the transversal one. The saturation curves $\beta\,\equiv\,\beta_0(\xi)$ are 
also plotted and labelled by the values of the normalized luminosity 
$\Lambda$ (for $\Lambda=\Lambda_{\rm{}crit}$ the saturation velocity
vanishes at infinity). Right: several examples of possible trajectories of a warm
electron cloud (thick curves). The initial condition is $\beta(t_0)=0$;
the values of $\xi(t_0)$ are indicated by a circle. A monoenergetic
distribution of scattering electrons was assumed,
$f(\gamma_\mathrm{e})=\delta({\gamma_\mathrm{e}-\bge})$. Cold cloud
trajectories are shown for the sake of comparison (thin solid curves).
The saturation curve is also plotted (dashed), corresponding to the 
luminosity parameter $\Lambda=10\mu$.}
\label{fig:acc-crit}
\end{figure*}

\section{Polarization from scattering on warm clouds near a black hole}
So far we assumed the cloud to be cold and we did not consider any
microscopic (random) motion of the scattering electrons within the cloud
volume. Hereafter, in order to account for the electron temperature we
average the equations (\ref{eq:scattI})--(\ref{eq:scattU}) over the
electron distribution $n(\vec{\beta}_\el)=n_\el\,f(\gamma_\el)$ in the
cloud comoving frame (CCF), where the electron distribution is entirely
isotropic and $\gamma_\el\,\equiv\,(1-\beta_\el^2)^{-1/2}$ is the
corresponding Lorentz factor. Stokes parameters are first evaluated in
the local electron comoving frame (ECF), Lorentz transformed to the CCF, 
and the result is averaged over random velocities of the electrons. We
obtain
\begin{eqnarray}
 \bar{I} &=& (1+\mathcal{A})\left(\bar{E}+\bar{P}^{ZZ}\right)+\mathcal{B}
 \left(\bar{E}-3\bar{P}^{ZZ}\right)-2\mathcal{A}\bar{F}^{Z},
  \label{eq:ics-warmI} \\
 \bar{Q} &=& \bar{P}^{YY} - \bar{P}^{XX},
 \label{eq:ics-warmQ} \\
 \bar{U} &=& -2\bar{P}^{XY},
 \label{eq:ics-warmU}
\end{eqnarray}
where $F^Z$ is the radiation flux component measured in direction of the
photon after scattering. We introduced mean values
\begin{equation}
 \mathcal{A}\equiv\textstyle{\frac{4}{3}}\left\langle\gamma_\el^2\beta_\el^2
 \right\rangle,
  \quad
 \mathcal{B}\equiv 1-\left\langle
 \frac{\ln[\gamma_\el(1+\beta_\el)]}{\beta_\el\gamma_\el^2}\right\rangle;
\label{eq:ics-hot-AB}
\end{equation}
$\langle\,x\,\rangle\equiv\int x\,f(\gamma_\el)\,\dd\gamma_\el$ denotes
averaging over the electron energy distribution; see Hor\'ak
(\yearcite{hor05b}) for details of the derivation of eqs.\
(\ref{eq:ics-warmI})--(\ref{eq:ics-warmU}).

The Lorentz boost from CCF to the static frame introduces the familiar
factor $\mathcal{D}^4$, where
$\mathcal{D}=\gamma^{-1}(1-\beta\cos\vartheta)^{-1}$ is the Doppler
factor, $\gamma=(1-\beta^2)^{-1/2}$, $\beta$ is the cloud bulk velocity 
and $\vartheta$ is the angle between this velocity and the scattered
photon direction:
\begin{equation}
I=\mathcal{D}^4\bar{I},\quad
Q=\mathcal{D}^4\bar{Q},\quad
U=\mathcal{D}^4\bar{U}.
\end{equation}
In this way the frequency-integrated Stokes parameters are found in
terms of the radiation stress tensor components. The relation between
the CCF components ($\bar{E},\bar{F}^A,\bar{P}^{AB}$) and the static
frame components ($E,F^a,P^{ab}$) is
\begin{eqnarray}
 \bar{P}^{X\!X\!} &=& \mathcal{D}^2\left[P^{xx}\left(\cos\vartheta-\beta\right)^2
 +\gamma^2\Delta\sin^2\vartheta\right], 
 \label{eq:pxx}
 \\
 \bar{P}^{ZZ} &=& \mathcal{D}^2\left[P^{xx}\sin^2\vartheta
 +\gamma^2\Delta\left(\cos\vartheta-\beta\right)^2\right], 
 \\
 \bar{P}^{YY} &=&  \bar{P}^{yy},
 \\
 \bar{E}~ &=& \gamma^2\left(E-2\beta F^z +\beta^2 P^{zz}\right),
 \label{eq:ptt}
 \\
 \bar{F}^Z &=& \mathcal{D}\gamma^2\,(\cos\theta - \beta)\,W.
\end{eqnarray}
where we denoted $\Delta=\beta^2 E-2\beta F^z+P^{zz}$ and
$W=(1+\beta^2)F^z - \beta\left(E + P^{zz}\right)$. Axial symmetry ensures
that $\bar{P}^{XY}$ vanishes for the primary radiation field originating from
the disk and impinging on the cloud.

Properties of the scattered signal in the static frame of course depend 
on the motion of the scatterer, which itself results from the interplay of 
acceleration mechanisms acting on the cloud. In our model the total 
four-force $f^\alpha$ is a superposition of the radiation and inertial 
terms. The radiation term is
\begin{equation}
f^\alpha =-\st\,n_\el\,\delta\mathcal{V}\,
\Big[\mathcal{C} T^{\alpha\beta}u_\beta - (\mathcal{A}+\mathcal{C})\,
T^{\rho\sigma} u_\rho u_\sigma u^\alpha\Big],
\end{equation}
where $\delta\mathcal{V}$ is the cloud volume,
$\mathcal{C}\,\equiv\,1+\frac{1}{2}\mathcal{A}$. The change of the cloud
internal energy by cooling is
\begin{equation}
\der{\bge}{s}=-\gamma\mathcal{A}\Lambda\mu\,
\Big(E_\star - 2\beta F_\star^{z} + \beta^2 P_\star^{zz}\Big),
\label{eq:acc-eomg3}
\end{equation}
where we expressed proper time $s$ in units of $\rs/c$ and we denote
$\mu\,\equiv\,m_\mathrm{p}/m_\mathrm{e}\simeq1$ (for electron-proton
plasma) or $\mu\simeq10^3$ (for electron-positron plasma). The
parameterisation by $\mu$ allows us to consider different types of
plasma. However, we note that the radiation field of a standard disk
alone is unable to efficiently accelerate protons, and so the case of
electron-positron plasma seems to be a more relevant application here
(\cite{fuk05}).

Dynamics of the
cloud as a whole is governed by equation of motion in the form
\begin{eqnarray}
 \der{\beta}{s}&=&\bge^{-1}\mathcal{C}\Lambda\mu
 \Big[\left(1+\beta^2\right)F_\star^{z}
  \nonumber \\
&& - \beta\,\left(E_\star+P_\star^{zz}\right)\Big] -
 \frac{\rs}{2\gamma z^2\xi^{1/2}},
\label{eq:acc-eomb3}
\end{eqnarray}
which includes also the black hole gravity. The last two equations couple the
cloud trajectory with the evolution of its internal temperature. In
order to close the set of equations we need to assume a specific form of
the electron energy distribution $f(\gamma_\mathrm{e})$.

\begin{figure*}[tb!]
\begin{center}
 \FigureFile(0.49\textwidth,0.49\textwidth){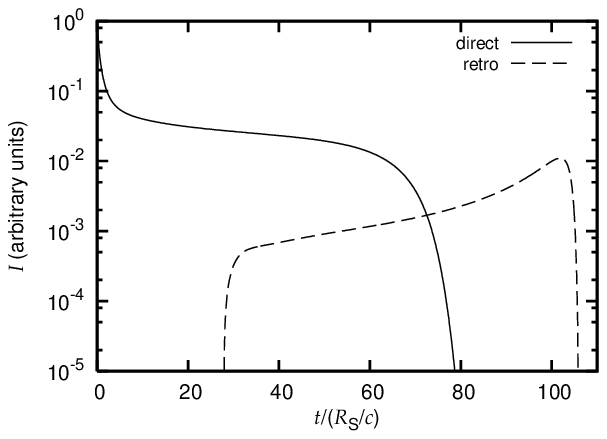}
 \hfill
 \FigureFile(0.49\textwidth,0.49\textwidth){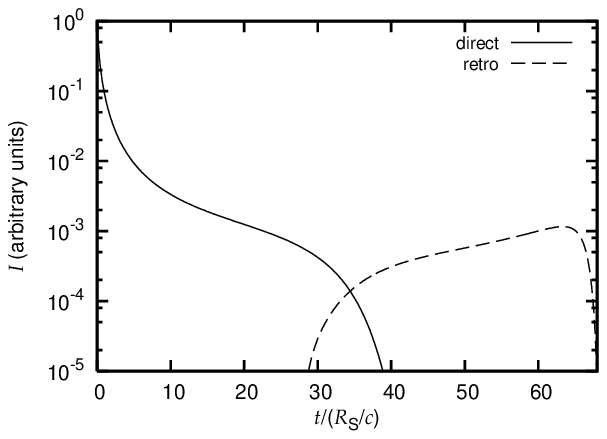}
 \FigureFile(0.49\textwidth,0.49\textwidth){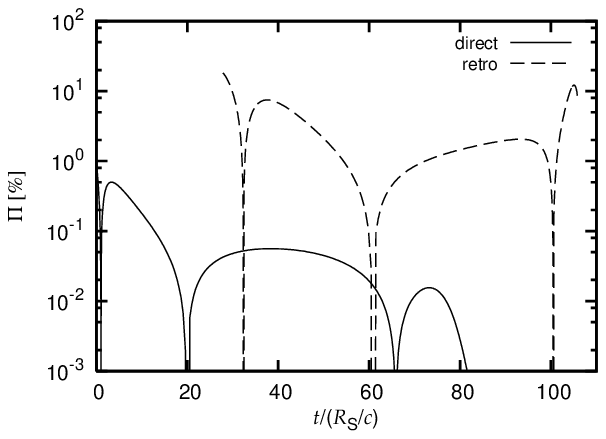}
 \hfill
 \FigureFile(0.49\textwidth,0.49\textwidth){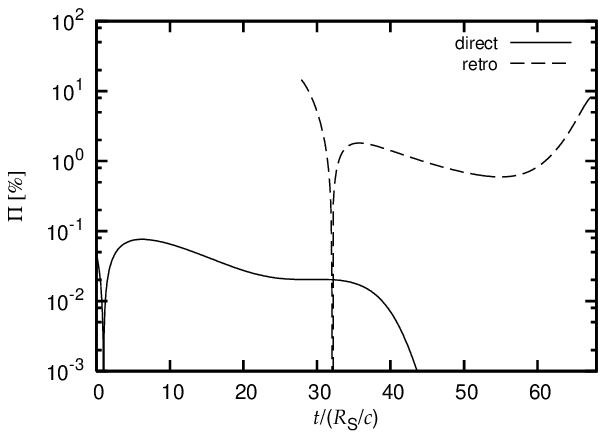}
 \FigureFile(0.49\textwidth,0.49\textwidth){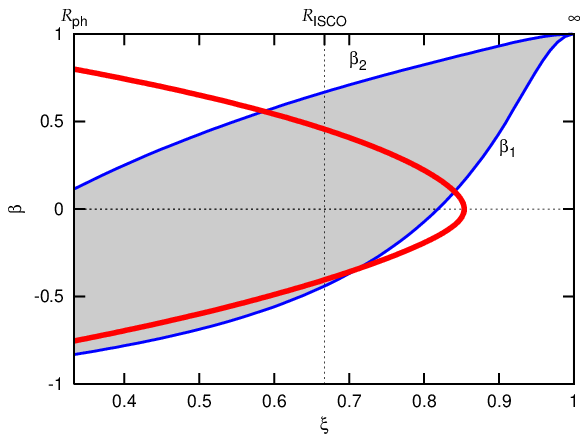}
 \hfill
 \FigureFile(0.49\textwidth,0.49\textwidth){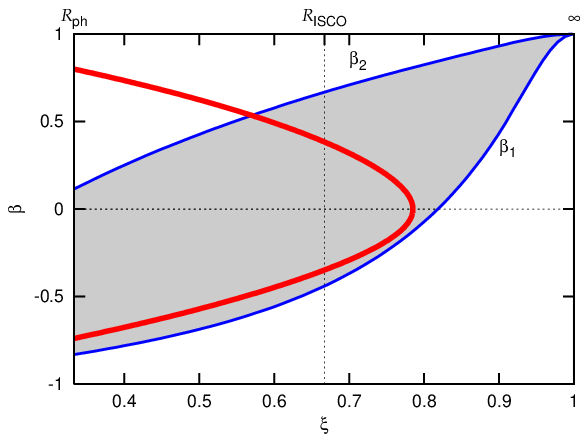}
\end{center}
\caption{Comparison between two typical cases with the identical initial
conditions except for the cloud temperature: a cold cloud (left panels,
$\bge=1$) versus a warm cloud (right panels, $\bge=3$ at start).
Examples are shown of the time behaviour of the observed intensity (top
row) and the magnitude of polarization (middle row). Contributions of
the retro-lensing images have been summed together and plotted (by a
dashed line); they are clearly distinguished from the signal produced by
the direct-image photons (solid line). The retro-lensing photons arrive
with a certain time delay with respect to the direct photons. The delay
is caused by photons taking different path and circling at $r\simeq\rph$ 
(the effect can be recognized by comparing the relative time shift of
the features). We also show, in the bottom row,
the corresponding velocity profile $\beta(\xi)$ of the cloud motion. The 
temperature evolution is given by solving eq.~(\ref{eq:acc-eomg3})
along the trajectory: the warm
cloud cools down to $\bge=1.3$ along its entire track.
Polarization vanishes at the moment when the cloud crosses one of the
curves $\beta_1(\xi)$, $\beta_2(\xi)$. In both cases the view angle was
$i=5$~deg, the disk luminosity $\Lambda=2$. The initial distance
$z(t_0)=\rph$ (i.e.\ $\xi=\frac{1}{3}$), $\beta(t_0)=0.8$.}
 \label{fig:acc-low-in}
\end{figure*}

In the case of cold clouds ($\mathcal{A}=0$, $\mathcal{C}=1$) the
dynamics is characterized by the saturation velocity $\beta_0(\xi)$ to
which particles are asymptotically  accelerated if inertial effects are
small compared to the radiation force (\cite{sik81}; \cite{ick89}). This
condition  corresponds to the limit of $\Lambda\rightarrow\infty$, 
${\dd\beta/\dd}s=0$. We thus obtain
\begin{equation}
\beta(z)\rightarrow\beta_0=\sigma-\sqrt{\sigma^2
 -1}\,,\quad
\sigma\equiv\frac{E_\star+F_\star^{z}}{2F_\star^{z}}\,.
\end{equation}  
Further, for polarization there are two critical velocities, 
$\beta_1(\xi)$ and $\beta_2(\xi)$, at which the polarization vector
changes its orientation between transversal and longitudinal one (this
can be seen from the condition $\bar{Q}=0$). Similar effect of
polarization direction changing with velocity of the scattering medium
was studied by Beloborodov (\yearcite{bel98}; for scattering in fast
outflows from accretion disks) and Hor\'ak \& Karas (\yearcite{hor05};
for radially moving clouds near a compact star). This behaviour is
demonstrated in the left panel of figure~\ref{fig:acc-crit}. The right
panel shows the corresponding influence of the radiation cooling of the
cloud on its bulk motion.

From eq.~(\ref{eq:acc-eomb3}) we find that the motion of warm clouds is
governed by the same equation as cold ones, provided that the normalized
luminosity is rescaled: 
\begin{equation}
\Lambda\rightarrow\Lambda^\prime\equiv\bge^{-1}\mathcal{C}\Lambda.
\end{equation}
In other words, by keeping the disk luminosity at a constant value of
the Eddington parameter, the cooling (\ref{eq:acc-eomg3}) gives rise to
trajectories gradually deviating from the corresponding cold-cloud
limit. 

The effect of changing temperature is such that the saturation velocity 
curve gradually moves across ($\beta,\xi$)-plane. This helps us to
understand the form of trajectories, which create loops (see 
fig.~\ref{fig:acc-crit}, right panel). The critical point is given by
condition $\beta=0$, $0<\xi<\infty$ and it defines the distance where
equilibrium between radiation and gravity can be reached. This is a
saddle type point, characteristic to an unstable equilibrium, which
makes the system behaviour qualitatively different from the case with a
central star as a source of primary irradiation (cp.\ with the critical
points examined in \cite{abr90}). Our result here resembles the case of
clouds with non-constant mass, discussed by Keane et al.\
(\yearcite{kea01}).

Now we can find the Stokes parameters of scattered light along the cloud
path. When determining the temporal evolution of observed intensity and
polarization we consider first three images in the observed radiation --
the direct (zeroth-order) image and two first-order images. The latter
are retro-lensing images formed by rays making a single round about the
black hole by the angle $\Phi=2\pi{\pm}i$, where $i$ is observer
inclination (we denote the retro-lensing images by $\pm$ signs to indicate
that they wind up in the opposite directions around the black hole). 
The higher-order images give a progressively weaker contribution to the
final signal, hence we can safely neglect them. The retro-lensing
photons give rise to peaks in the observed signal occurring with a
characteristic mutual time lag after the direct-image photons. Duration
of these features is very short -- comparable to the light crossing time
-- and the time span of the plots can be therefore scaled by
$t_{\rm{}lc}\simeq1.5\times10^{-4}M/(10M_\odot)\;\mbox{[sec]}$.

Typical profiles of the resulting lightcurves and polarization curves
are shown in figure~\ref{fig:acc-low-in}. Here we notice the difference
in the lightcurve profiles and the the polarization curves, which is
caused by effect of the cloud initial temperature and its gradual
cooling.  The contribution of the retro-lensing images to the total
observed flux is maximum when the scatterer, the black hole and the
observer are well-aligned ($i\rightarrow0$) and the scatterer moves
towards black hole ($\beta\rightarrow-1$). In this situation a
non-negligible fraction of photons are scattered toward on the light
circular photon orbit, their energy is Doppler boosted, and eventually
they are redirected toward the observer. On the other hand, polarization
magnitude attains maximum in a slightly different direction whereas, for
symmetry reasons, $\Pi$ vanishes for a  strictly aligned observer. The
direct image photons are of course most intense in the opposite case
($\beta\rightarrow1$), and so there is an interplay of different orders
of the images sensitive to the model parameters.

Identifying the short-duration features in lightcurves could
help confirming the existence of the photon circular orbit around the
central body, for which black holes are the most likely cause.
Polarimetric resolution provides additional information about the
source and it could be used to constrain the parameters of the
black hole. 

\section{Discussion and conclusions} 
In X-rays, future polarimeters could be employed to probe jets and winds
in strong gravitational fields of the central compact object.
Polarimetry is a powerful tool that can provide additional information, 
which would be difficult to obtain by other techniques such as
traditional photometry and spectroscopy. In this way  polarimetry helps
to discriminate between different geometries and physical states of
sources where accretion processes are accompanied by fast radial motion
of the blobs of material. Naturally, this goal would require sufficient
sensitivity in X-rays; the scattered signal is mixed with primary
photons, which reduce the final polarization. We have seen that the
predicted features are flashing for only a brief period of time and the
maximum polarization degree is typically $\simeq1$ percent or less. 

In order to allow the analytical treatment we employed various
simplifications: we considered the bolometric quantities and the Thomson
cross-section for scattering of primary photons on electrons (rather
than Compton scattering and the Klein-Nishina cross-section; see
\cite{mel89}; \cite{ski94}; \cite{mad00}). We also assumed that the flow
is not magnetically dominated, although astrophysically realistic models
require magnetohydrodynamic effects to be taken into account
(\cite{beg84}; \cite{bes04}). Effects of general relativity were taken
into account in the limit of the non-rotating black hole spacetime (we
neglected the effects of frame-dragging for the sake of simplicity; cf.\
\cite{vok91}). Likewise, we adopted the simplest possible parameterization
of the disk emissivity via the standard Shakura--Sunyaev model; 
this could be improved by including relativistic effects (\cite{pag74})
and a more realistic description of the disc itself. These changes will
be necessary to provide quantitative and astrophysically realistic results
for the polarization degree, however, we do not expect any qualitative
change regarding the signature of indirect photons. The predicted 
polarization is either parallel or perpendicular to the projection of 
cloud velocity onto the observing plane. 

\medskip
We gratefully acknowledge fruitful discussions at the Institute  of
Theoretical Physics in Prague and we thank for helpful comments that we
have received from the referee. We also acknowledge the financial
support from the Academy of Sciences (ref.\ IAA\,300030510) and from
the Czech Science Foundation (ref.\ 205/03/H144). The Astronomical
Institute has been operated under the project AV0Z10030501.

\end{document}